# On the nature of dark matter in the Coma Cluster

K. Zioutas[1], D.H.H. Hoffmann[2], K. Dennerl[3], T. Papaevangelou[4]

[1] University of Patras, 26504 Patras, Greece, and, CERN, 1211 Geneva 23, Switzerland.   Email: zioutas@cern.ch

[2] Institut für Kernphysik, TU-Darmstadt, Schlossgartenstr. 9, 64289 Darmstadt, Germany.   Email: hoffmann@physik.tu-darmstadt.de

[3] Max-Planck-Institut für extraterrestrische Physik, Giessenbachstraße, 85748 Garching, Germany.  Email:  kod@mpe.mpg.de

[4] IRFU, Centre d'Études Nuclaires de Saclay, Gif-sur-Yvette, France.
Email:  thomas.papaevangelou@cea.fr

**Abstract:**

Recent precise observations of the 2.7 K Cosmic microwave Background Radiation (CBR) by the *Planck* mission toward the Coma cluster are not in agreement with X-ray measurements. To reconcile both types of measuring techniques we suggest that unstable dark matter is the cause of this mismatch. Decaying dark matter, which gravitationally dominates the galaxy cluster, can affect the estimated hot plasma content, which is then missing in the measured SZ effect from exactly the same place in the sky. The model independent lifetime of dark matter decaying entirely to X-rays is estimated to be about $6 \cdot 10^{24}$ sec; this lifetime scales down with the fraction of the radiatively decaying dark matter. In addition, it is shown that the potential of such dark matter investigations in space is superior to the largest volume Earth-bound dark matter decay searches. Other clusters might provide additional evidence for or against this suggestion.

## 1. Introduction

Galaxy clusters are the most massive gravitationally bound structures in the Universe, with an accumulated matter up to about $10^{15} M_\odot$ [1]. They are considered to consist mostly of dark matter, whose nature still remains a big mystery. Therefore, clusters are highly important sites to test fundamental cosmological parameters [2]. After all, it was the Coma cluster ($M_{coma} \approx 7 \cdot 10^{14} M_\odot$), where Zwicky in 1933 [3] concluded on the existence of "*dunkle Materie*". In this work we argue that an additional X-ray emission (from dark matter decays) could be at the origin of the derived mismatch between the results obtained from two observations of the Coma cluster based on different processes, due to: a) The Compton scattering in the meV range (2.7 K) off the hot electron cloud in the intercluster medium (ICM), and b) The Bremsstrahlung emission in the keV range (hard X-rays) due to electron scattering by the same hot plasma. The observed photon energies are separated by several orders of magnitude. Then, the hot Coma cluster at 102 Mpc remains a target of opportunity for dark matter investigations. As it was pointed out in ref. [4], such decay X-rays (with a broad energy distribution) would resemble that of a "*ghost hot plasma*". Independent measurements, for example, with the celebrated Sunyaev-Zeldovich (SZ) effect which is sensitive to the electron content of the intercluster medium (see below), can provide a cross check of the widely assumed hot plasma scenario as the only source for the X-ray luminous cluster. In ref. [4] it was mentioned that Galaxy clusters require additional gas physics, and some as yet unknown non-gravitational processes. In particular, it was predicted that upcoming precise measurements of the SZ effect, might result in discrepancies when compared to expectations from X-ray observations. It is this aspect, which this work addresses following recent measurements targeting the Coma cluster. In fact, an unstable dark matter component, whose products are not taken into account, can result in discrepancies between otherwise related methods; because, it can manipulate the derived cluster plasma properties in an unpredictable way, as long as the spectral shape and amplitude of non-thermally created X-rays, e.g. from particles decays, are unknown.

## 2. **The Coma cluster observations**

About 90% of the Coma cluster mass is dark matter, for which it is widely accepted so far: "*the vast majority of the cluster masses appears to be dark and does not emit any detectable radiation*" (see e.g. ref.[1]). Though, exactly the same places are highly X-ray bright. We suggest here that massive Galaxy clusters are of potential interest in dark matter research, since the conventional X-ray emission (i.e., Bremsstrahlung from the ordinary hot ICM) might well include an as yet unidentified X-ray contribution caused by dark matter decay. To decipher from a cluster observation any property of its dark matter content, poses a challenge of utmost importance.

This work was triggered by the recent precisely measured SZ effect from the Coma cluster by the *Planck* mission [5]. Figure 1 compares the SZ data (by *Planck* and WMAP) with the expected distribution following X-ray measurements with the ROSAT and XMM-Newton observatories (the 2σ uncertainties are indicated by the shaded boxes) [6]. The discrepancies are largest at the center (θ < 10 arcmin). The different level of the expected SZ effect by each X-ray measurement fits actually the reasoning of this work: ROSAT has a detection sensitivity in the ~0.1 – 2 keV range, while XMM-Newton covers a much wider energy range (~0.1 – 10 keV). The (unknown) spectral shape and intensity of decay X-rays from unstable dark matter particles, relative to the thermal X-ray spectrum of the hot ICM, affects the conclusions derived for the electron density and temperature profile of the radiating plasma; this is used in estimating the expected Compton scattering of the 2.7K CBR that is underlying the SZ effect. We recall that the Bremsstrahlung emission from the hot plasma is $\mathbf{L_x} \propto \rho^2 \cdot T^{1/2}$, while $\mathbf{\Delta T_{SZ}} \propto \rho \cdot T$ gives the strength of the SZ effect (density = ρ and temperature = T) [6]. The discrepancy between the measured and expected SZ effect shows a *deficit* in the SZ effect by about 20% (see Figure 1). This discrepancy could be evidence for an underestimation of the available "SZ electrons" or an overestimation of the product $\rho^2 \cdot T^{1/2}$. However, observationally, and assuming that the only source of the X-rays is Bremsstrahlung emission, the temperature of the hot ICM of the Coma cluster is known with ~1% precision: kT= (8.25±0.1) keV [7], excluding thus that the mentioned overestimation by as much as 10-20% can be due to a wrong temperature. Because, in addition, an uncertainty of 1% in temperature implies a similar variation in density, and this in the opposite direction: $\delta\rho/\rho \sim -\delta T/T$ [8]. An additional component of X-rays coming from dark matter decays can affect not only the derived plasma density but also its apparent temperature. Fortunately, the SZ effect can cross check independently the same plasma properties. The observed disagreement points to a new diagnosis potential when combing SZ effect and X-ray measurements. Then, to reconcile the discrepancies shown in Figure 1 via the relevant relations on $L_x$ and $\Delta T_{SZ}$ given before, the electron density ($\rho$) is the only remaining "free" parameter. In other words, the SZ deficit implies a decreased electron density at the 15-20% level. Taking into account the different expectations in Figure 1, we use in the following approximate estimation a conservative density deficit of ~10%. However, this direct measurement of the electron content and its temperature remains in tension with the expectations following two previous X-rays observations by the ROSAT and the XMM-Newton orbiting telescopes [6]. So far, the origin of the enormous X-ray emission, $L_x \approx 5.1 \cdot 10^{44}$ erg / s / (2-10 keV) [9] is taken to be an ~8 keV hot plasma, which is supposed to constitute the 'visible' part of the ICM. We recall, however, that this hot plasma resides inside the gravitational potential well provided by the dominating dark matter there.

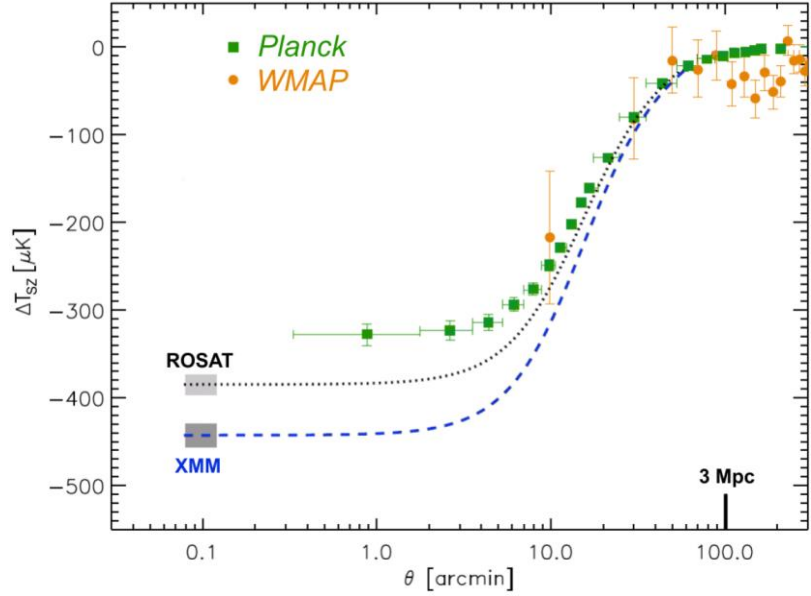

**Figure 1** *Profile of the SZ effect toward the Coma cluster [5]. The expected distributions following the results by the X-ray observatories ROSAT and XMM-Newton demonstrate the deficit seen by the recent SZ measurements by Planck [6]. This figure was reproduced by permission of the AAS.*

Note that X-ray observatories have large effective detector area, and, a field of view that is comparable in size with that of the relatively nearby Coma cluster. The highly telescopic performance combined with the enormous dark matter content there, results to a figure of merit superior to direct Earth-bound searches for unstable dark matter.

**Numerical examples**

**a)** Figure of merit (FOM): for an order of magnitude estimate, we use for the size of the Coma cluster R = 1Mpc and its dark matter being equal to $6 \times 10^{14} M_{solar} \approx 6 \times 10^{71}$ GeV/c$^2$. This defines a detection efficiency of a distant X-ray source at the Coma cluster (D = 102 Mpc ≈ $3 \cdot 10^{26}$ cm) of about $10^{-51}$, following the $1/4\pi D^2$ law for an effective area taken to be equal to

1000 cm$^2$ (e.g., XMM-Newton). Further, the background is neglected, since the X-ray signal from the Coma cluster is quite strong. Then, a figure of merit can be defined as: FOM= 6x10$^{71}$GeV/c$^2$·10$^{-51}$ =: 6x10$^{20}$. Similarly, for an earth bound detector, say, of 1000 m$^3$ sensitive volume and 100% detection efficiency, assuming 0.3 GeV/cm$^3$ as the local dark matter density, its total dark matter content is 3x10$^8$ GeV/c$^2$; the corresponding value for its FOM becomes equal to 3x10$^8$. Then, interestingly, the FOM for the orbiting X-ray telescope observing the Coma cluster is superior to the largest detectors by a factor of about 10$^{12}$. This simple comparison shows the enormous potential an orbiting X-ray observatory can have (combined with the SZ effect), in detecting dark matter decaying to X-rays in cluster. Of course, galaxy cluster observations inherit a physical background, which is associated with the hot plasma (ICM) from the same region. Therefore, for decaying dark matter its optimum signature in space might depend on the spectral shape and the amplitude of the decay X-rays compared to that of the Bremsstrahlung of the underlying hot ICM. In other words, an additional contribution from radiatively decaying dark matter can eventually falsify the underlying thermal spectrum depending on their rest mass or the production mechanism. For this, we mention the massive axion spectra derived in ref's [10,11,13] as relevant examples. To be more specific, if dark matter related X-rays peak at, say, 8 keV, then they may affect the spectrum of a few keV hot ICM stronger than an ICM at 8 keV or even hotter. This makes additional, but precise, SZ investigations with more galaxy clusters, with well known thermal properties, of potential interest.

**b)** Lifetime: for a ~10% missing electron content of the hot ICM, the corresponding X-ray luminosity which is not Bremsstrahlung in origin, is ~20% (due to the $\rho^2$ – dependence). I.e., the component of the X-ray luminosity coming from the assumed decaying dark matter constituents is: $L_{x-DM}$≈0.2x5.1·10$^{44}$ erg/s ≈10$^{44}$ erg/s. Assuming further that ~90% of the Coma cluster mass is dark matter, i.e., $M_{DM}$ ≈ 6·10$^{14}$ M$_\odot$, we arrive, after simple conversions, to a model independent mean lifetime (τ) of the unstable dark matter in the Coma cluster of τ≈$M_{DM}/L_{x-DM}$ = … ≈ 6·10$^{24}$ sec ≈ 10$^6$x the present age of the Universe.

## 3. Discussion

Inspired by the massive axions of the Kaluza-Klein type [4,10,11], we suggest that the radiative decay of such or any other unstable trapped massive exotica (see, for example, ref's [12,13]) might constitute (partly) the dark matter budget of the Coma cluster, or other clusters too. Hence, such particle constituents can give rise to a misinterpretation of the origin of the cluster X-ray luminosity. In this work, we argue that unstable dark matter that holds gravitationally, e.g. the Coma cluster, can reconcile the discrepancies between the expectations from both X-rays measurements and the recent precise measurement of the SZ effect from the same place in the sky. It is worth mentioning that still, X-ray observations are used as one of the fundamental methods to recover the galaxy cluster mass [14], and of course, the temperature of the ICM. This does not hold, if additional as yet hidden non-thermal X-ray sources are being ignored. Decaying dark matter is one option addressed here. Due to the enormous quantity of dark matter accumulated at those places in the Universe the unstable dark matter can be, for example, very long lived radiatively decaying particles like massive axions or other candidates [10-13]. Extending such investigations with additional Galaxy clusters at various distances, with different temperatures of their ICM, different age, etc., will be important to establish the scenario presented in this work, or instead any other alternative ones (see e.g. ref. [6]).

Furthermore, it is shown that targeting dark matter agglomerates as the accumulated ones in galaxy clusters, the potential of such dark matter investigations in space surpasses the best Earth-bound direct experimental searches with the largest volume detectors.

## 4. Conclusions

Whether dark matter is (un)stable, is not proven yet. It is argued in this work that radiatively decaying but long lived dark matter constituents could be at the origin of the observed discrepancies between related measurements toward the Coma cluster. A combined evaluation of the results derived from both kind of measuring techniques point at an additional non-thermal X-ray source in the ICM, whose ignorance can be the cause for the mismatch SZ effect *versus* Bremsstrahlung. Of course, assuming that CBR photons do not experience some kind of up-scattering by DM as it happens with the hot ICM; though, this possibility cannot be excluded, and, it is within the exotic scenario of this work. As it was predicted long time ago, the observed SZ deficit fits the scenario suggested in this work: the extra X-rays coming from unstable dark matter, from exactly the same place in the sky, overestimate the mass of the actual plasma (ICM), affecting thus the expectations for the SZ effect. I.e., there are less electrons there than anticipated by the spectral shape and intensity of the X-rays. The two unequal expectations by ROSAT and XMM-Newton following their measured X-rays might fit the reasoning of this work, because of the different energy range covered by each X-ray observatory. The derived mean lifetime, assuming that all the dark matter in the Coma cluster undergoes spontaneous radiative decays, is $\tau \approx 6 \cdot 10^{24}$ sec. This result is model independent, and, it scales down with the actual fraction (f) of unstable particles in the cluster. For example, if f=1%, then $\tau \approx 6 \cdot 10^{22}$ sec, which is of course still a long lifetime.

More investigations of this kind using a wide energy band-width of telescopes have the potential to provide more insight in the dark matter properties, which are essential for our understanding of the Universe. In any case, such additional and independent investigations might definitely establish, or reject, the scenario presented here, or, any other possible ones, e.g., the one proposed recently in ref. [6].

## Acknowledgments

We thank Andrea Lapi for permission to use Figure 1, and, Mary Tsagri for her real help while preparing this work.